\documentclass[a4paper,11pt]{article}
\usepackage{jcappub} 

\usepackage{float}

\usepackage{lineno}

\usepackage{etoolbox}
\newcommand*\patchAmsMathEnvironmentForLineno[1]{%
  \expandafter\let\csname old#1\expandafter\endcsname\csname #1\endcsname
  \expandafter\let\csname oldend#1\expandafter\endcsname\csname end#1\endcsname
  \renewenvironment{#1}%
    {\linenomath\csname old#1\endcsname}%
    {\csname oldend#1\endcsname\endlinenomath}%
}
\newcommand*\patchBothAmsMathEnvironmentsForLineno[1]{%
  \patchAmsMathEnvironmentForLineno{#1}%
  \patchAmsMathEnvironmentForLineno{#1*}%
}
\patchBothAmsMathEnvironmentsForLineno{equation}
\patchBothAmsMathEnvironmentsForLineno{align}
\patchBothAmsMathEnvironmentsForLineno{gather}
\patchBothAmsMathEnvironmentsForLineno{multline}
\patchBothAmsMathEnvironmentsForLineno{flalign}

\title{\boldmath Oscillations of the black hole photon ring as a probe of ultralight dilaton fields}

\author[a]{Chunlong Li,}
\author[b]{Chao Chen,}
\author[b]{Xiao Yan Chew}

\affiliation[a]{School of Physics, Hefei University of Technology, Hefei, Anhui 230009, China}
\affiliation[b]{School of Science, Jiangsu University of Science and Technology, Zhenjiang, 212100, China}

\emailAdd{chunlong1024@hfut.edu.cn}
\emailAdd{cchao012@just.edu.cn}
\emailAdd{xiao.yan.chew@just.edu.cn}

\abstract{Advancements of very long baseline interferometry (VLBI) have facilitated unprecedented probing of superradiant phenomena in the vicinities of supermassive black holes (SMBHs), establishing an ideal laboratory to detect ultralight bosons beyond the Standard Model. In this study, we delve into how ultralight dilaton clouds, formed via SMBH superradiance, impact the black hole photon rings. Our focus is on the dilaton-electromagnetic coupling term of the form $f(\phi)F_{\mu\nu}F^{\mu\nu}$. By integrating geometric optics with plasma refractive effects in accretion environments, we demonstrate that the dilaton cloud dynamically alters the plasma frequency. Through systematic ray-tracing simulations covering a range of photon frequencies and dilaton coupling strengths, we reveal a photon ring oscillation that follows the period of that of the dilaton field. As the dilaton mass increases, this oscillation becomes suppressed due to the washout effect of the dilaton-induced correction term over the light path integration. We further evaluated the observability of such dilaton-induced photon ring oscillations with current radio interferometric capabilities. Our estimates indicate that this effect could potentially constrain the dilaton-photon coupling to $g_{\phi\gamma}\lesssim 10^{-11}\text{GeV}^{-1}$ for dilaton masses $\mu \lesssim 10^{-18}\,\mathrm{eV}$.}

\begin{document}
\maketitle
\flushbottom

\section{Introduction}
\label{sec:intro}

Taking advantage of the Very Long Baseline Interferometer (VLBI) technology, the Event Horizon Telescope (EHT) opens a new era of probing physics under extreme conditions near the event horizon of a supermassive black hole (SMBH)~\cite{EHT2019dse,EHT2019ths,EHT2019ggy,EHT2019pgp}. The observed images confirm the Kerr hypothesis in general relativity, not only providing strong evidence for their existence, but also offering a golden opportunity to explore exciting phenomena in the regime of strong gravity. At the center of the galaxy M87, the compact radio source is resolved as an asymmetric, bright emission ring surrounding a central dark region. Although many details remain under active investigation, theoretical studies suggest the presence of a strongly lensed structure—referred to as the “photon ring”—underlying the dominant direct emission from the accretion flow \cite{Gralla:2019xty, Narayan:2019imo}. With the help of sufficient high-resolution imaging, complexities from astrophysical effects can be mitigated, as the size and shape of the photon ring are totally determined by the instabilities of photon orbits predicted by geodesic equations, which makes the the black hole photon ring potential probe to test the effect of strong gravity and related new physics \cite{Johnson:2019ljv, Gralla:2020nwp, Gralla:2020yvo, Gralla:2020srx, Gralla:2020pra,Hou:2021okc,Zhang:2024lsf,Hou:2022eev}.

In addition to testing gravity, horizon-scale observations also offer a promising avenue for probing particle physics, particularly in the search for ultralight bosons. These particles naturally arise in theories involving extra dimensions \cite{Arvanitaki:2009fg} and are compelling candidates for dark matter \cite{Preskill:1982cy,Abbott:1982af,Dine:1982ah,Sakharov:1996xg}. If their mass lies within an appropriate range, ultralight bosons can be efficiently accumulated around a rotating black hole via the superradiance mechanism \cite{Penrose:1971uk,1971JETPL..14..180Z,Press:1972zz,Damour:1976kh,Zouros:1979iw,Detweiler:1980uk,Strafuss:2004qc,Dolan:2007mj,Brito:2015oca}. The formation of the superradiant bosonic cloud can extract rotational energy from a black hole, leading to a reduction in its spin. This process may leave observable signatures in the black hole’s shadow or photon ring \cite{Roy:2019esk,Roy:2021uye,Chen:2022nbb,Chen:2022kzv,Saha:2022hcd}. Moreover, superradiant bosons may also give rise to detectable signals through non-gravitational couplings to Standard Model particles. For instance, the EHT polarimetric measurement on $\mathrm{M87}^\star$ has been proposed to search for the existence of the superradiance axion-like particles through the birefringence effects caused by their interaction with photons \cite{Chen:2019fsq, Carroll:1989vb,Harari:1992ea,Roy:2023rjk}. A coherently oscillating axion field will lead to a periodic modulation to the electric vector position angles of linearly polarized radiation from the accretion flow, enabling such polarimetric measurements to serve as a powerful tool for constraining the axion parameter space \cite{Chen:2021lvo,Chen:2022oad}.

The dilaton is an another hypothetical ultralight bosonic field, which plays a significant role in theoretical cosmology. In string theory-inspired models, the dilaton arises naturally as a scalar partner to the graviton, governing the strength of gravitational interactions through its coupling to spacetime geometry \cite{Damour:1994zq,Green:1987sp}. It has been proposed as a candidate for driving dynamical dark energy, offering a compelling explanation for the late-time accelerated expansion of the universe \cite{Wetterich:1987fm,Gasperini:2001pc}. A time-varying vacuum expectation value of the dilaton field could leave detectable imprints on the cosmic microwave background (CMB) anisotropies and influence the formation of large-scale structures \cite{Tocchini-Valentini:2001wmi,Davis:2005au}. Additionally, the dilaton may induce variations of fundamental constants, e.g., the fine-structure constant over cosmic time, which could be tested via astrophysical observations \cite{Bekenstein:2020wgp,Sandvik:2001rv,Stadnik:2015kia}. Furthermore, recent advances in quantum metrology have enabled stringent tests of dilaton-induced ‘fifth forces’ or deviations from the inverse-square law of gravity \cite{XIONG2024,Yin:2025uzf,Yin:2022geb,Fischer:2024eic}. Given its rich theoretical and phenomenological properties, it is imperative to expand the range of search strategies for the dilaton.

In this work, we focus on the dilaton cloud formed via the superradiant mechanism around supermassive black holes. Our analysis begins with the non-minimal coupling term $f(\phi)F_{\mu\nu}F^{\mu\nu}$, which describes the interaction between the dilaton and the electromagnetic field. Applying the geometric optics approximation, we derive the modified dispersion relation for photon propagation in the presence of the superradiant dilaton field, demonstrating that the dilaton field modifies the effective mass term of photons through its influence on the plasma frequency. Using ray-tracing numerical simulations, we explore how this modification affects the morphology of the black hole's photon ring, revealing periodic oscillations in its shape and size that correspond to the oscillation frequency of the superradiant dilaton field. Furthermore, within the framework of a pressureless perfect fluid and spherically symmetric accretion model, we estimate the magnitude of this effect, and assessed the constraint that current radio observational capabilities can place on the dilaton-photon coupling constant $g_{\phi\gamma}$.

The structure of this paper is organized as follows. In Section \ref{introdila}, we briefly review the superradiance mechanism of black holes. In Section \ref{ppdp}, we derive the modified dispersion relation and equations of motion for photons under the influence of the dilaton field. In Section \ref{sbr}, we employ the ray-tracing method to numerically simulate the resulting modifications in the photon ring morphology. In Section \ref{est}, we assess the observability of the superradiant dilaton field under current instrumental sensitivity thresholds. Section \ref{con} concludes with a summary and discussion of implications. Throughout this study, we work in units where $\hbar = c = 1$, and  adopt the metric convention $(-,+,+,+)$.

\section{Superradiant dilaton cloud surrounding black holes}
\label{introdila}

The superradiance mechanism is governed by the interplay between two characteristic physical scales. The first one is the boson's reduced Compton wavelength $\lambda_c$, which is related with the boson mass $\mu$ as $\lambda_c \equiv 1/\mu$. Another one is the gravitational radius $r_g \equiv G M$ with $M$ is the mass of the Kerr black hole. When $\lambda_c$ is comparable to $r_g$, the superradiance mechanism becomes most effective, leading to the exponential growth of a boson cloud around the black hole \cite{Penrose:1971uk,1971JETPL..14..180Z,Press:1972zz,Damour:1976kh,Zouros:1979iw,Detweiler:1980uk,Strafuss:2004qc,Dolan:2007mj,Brito:2015oca}. For supermassive black holes with masses between $10^6 \mathrm{M_\odot}$ to $10^{10} \mathrm{M_\odot}$, the corresponding boson mass $\mu$ falls within the range of $10^{-20}$eV to $10^{-16}$eV, well within the ultralight regime. Neglecting the self-interaction of the dilaton, its dynamics in a curved spacetime are governed by the Klein–Gordon equation, which takes the form:
\begin{equation}
    (\nabla^\mu \nabla_\mu - \mu^2) \phi = 0 ~.
    \label{kge}
\end{equation}
Here, $\nabla_{\mu}$ denotes the covariant derivative of $\phi$ in the background of Kerr metric with the angular momentum $J$ in Boyer-Lindquist (BL) coordinates $x^{\mu}=[t, r, \theta, \varphi]$. The metric of Kerr black hole in the BL coordinates is given by
\begin{equation}
 ds^2 = - \left(  1 - \frac{2 r_g r}{\Sigma}   \right) dt^2 + \frac{\Sigma}{\Delta} dr^2 + \Sigma d\theta^2 + \frac{A}{\Sigma} \sin^2 \theta d\varphi^2 - \frac{4 r_g a r}{\Sigma} \sin^2 \theta dt d\varphi \,,  
\end{equation}
where 
\begin{align}
 \Delta &= r^2 - 2 r_g r + a^2 \,, \\
 \Sigma &= r^2 + a^2 \cos^2 \theta \,, \\
 A &= (r^2+a^2)^2 - a^2 \Delta \sin^2 \theta \, ,
\end{align}
and $a$ is the dimensionless spin defined as $a=J/(M r_g)$. The Kerr black hole possesses two horizons, which are the outer horizon $r_+$ and inner horizon $r_-$. Their explicit form are $r_\pm=r_g\pm\sqrt{r_g^2-a^2}$ which can be obtained by directly solving $\Delta=0$. Eq.\,(\ref{kge}) is separable for the Kerr black hole when the scalar field $\phi$ reads \cite{Brill:1972xj,Carter:1968ks},
\begin{equation}
    \phi(t, r, \theta, \varphi)=e^{-i\omega t+im\varphi}R_{nlm}(r)S_{lm}(\theta) ~,
    \label{phiprofile}
\end{equation}
where $R_{nlm}(r)$ is the radial function and $S_{lm}(\theta)$ is the spheroidal harmonics. The substitution of Eq.~\eqref{phiprofile} into Eq.~\eqref{kge} yields the following radial part of Teukolsky equation after the separation of variables has been applied,
\begin{align}
    &\left[ \left( \left(r^2+a^2 \right)\omega_{nlm}-am \right)^2 - \left( \frac{r-r_+}{r-r_-}  \right) \left( a^2 \omega^2_{nlm} -2 ma\omega_{nlm} + \mu^2 r^2 + P_{lm} \right)  \right] R_{nlm} \nonumber \\
    &+\left(  \frac{r-r_+}{r-r_-} \right) \frac{d}{dr} \left[ \left( \frac{r-r_+}{r-r_-} \right)  \frac{d R_{nlm}}{dr}  \right] =0 ~, \label{rad_Teu}
\end{align}
where $P_{lm}$ and $\omega_{nlm}$ are the angular and energy eigenvalues, respectively. 
$(n,l,m)$ are quantum numbers satisfying $n \geq l+1, l \geq 0$ and $l \geq|m|$. The boundary condition of the wavefunction is the ingoing at the Kerr black hole's outer horizon and going to zero at infinity, which makes the eigen-frequencies $\omega_{nlm}$ generally take a complex form $\omega_{nlm}=\omega_{nlm}^r+i\omega_{nlm}^i$. The onset of superradiance can be clearly demonstrated by considering the regime where the dimensionless coupling parameter $\alpha\equiv r_g/\lambda_c$ satisfies
$\alpha \ll 0.1$ \cite{Ternov:1978gq,Detweiler:1980uk}. In this limit, the real part $\omega^r_{nlm}$ and the imaginary part $\omega_{nlm}^i$ can be expressed as
\begin{align}
    &\omega_{nlm}^r=\mu\left(1-\frac{\alpha^2}{2n^2}+\mathcal{O}(\alpha^4)\right) ~, 
    \\
    &\omega_{nlm}^i\propto\alpha^{4l+5}\left(m\Omega_H-\omega_{nlm}^r\right)\left(1+\mathcal{O}(\alpha)\right)
    \label{omegai}
\end{align}
with the definition $\Omega_{\rm H}\equiv a/(2r_+)$. The higher order terms of $\alpha$ can be found in \cite{Baumann:2019eav}, which contains the dependence on the quantum number $l$ and $m$. Note that $\omega_{nlm}^i$ in Eq.\,(\ref{omegai}) can be positive when the following condition is met for a particular set of quantum numbers $(n,l,m)$:
\begin{equation}
    \Omega_{H} > \frac{\omega^r_{nlm}}{m} ~.
    \label{src}
\end{equation}
This leads to an exponential growth of wave with the timescale $\tau_{SR}= 1/\omega_{nlm}^i$, i.e., the superradiance process.

Eq.\,(\ref{rad_Teu}) is solved numerically to obtain the solution for $R_{nlm}(r)$. The numerical calculations in \cite{Dolan:2007mj} demonstrates that
the state with the lowest energy state among the ones satisfying the superradiance condition has the highest superradiant rate, which corresponds to $n=2, l = 1, m = 1$ and $S_{11}(\theta)=\sin\theta$. In this work, we adopt this state as our benchmark, whose explicit form is given by:
\begin{align}
    \phi(t, r, \theta, \varphi)=\phi_{\text{max}} e^{-i\omega t+im\varphi}\mathcal{R}_{211}(r)\sin(\theta) ~,
\end{align}
where $\mathcal{R}_{211}$ is defined as $\mathcal{R}_{211}(r)\equiv R_{211}(r)/R_{211}(r_{\text{max}})$ with $r_{\text{max}}$ denoting the radius at which the radial profile $R(r)$ reaches its maximum, and $\phi_{\text{max}}$ is the corresponding maximum field value. The value of $r_{\text{max}}$ can be approximated as $r_{\rm max} \approx n^2/(2\alpha^2) r_g$ in the limit of $\alpha\ll 0.1$. We depict profiles of $\mathcal{R}_{211}(r)$ and its derivatives $\mathcal{R}'_{211}(r)$ with respect to $r$ in Fig.\,\ref{Kerr} for $a=0.99$ and several values of $\alpha$.


\begin{figure}[t]
    \centering
    \includegraphics[width=0.45\textwidth]{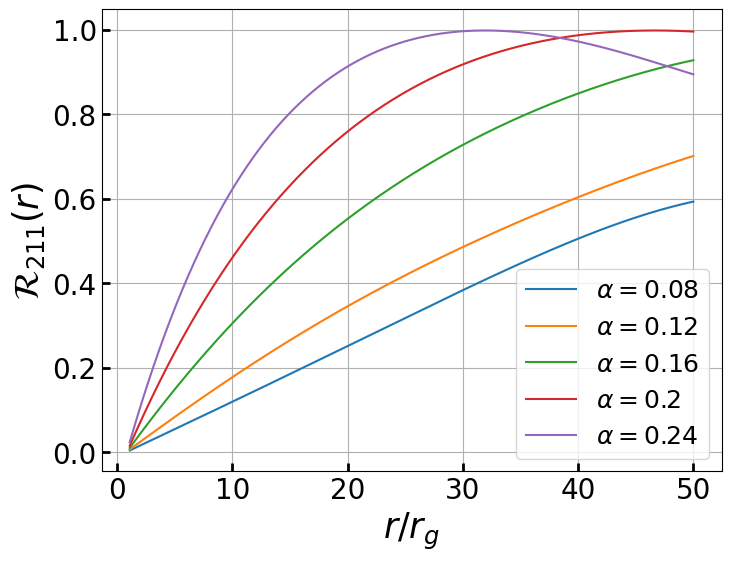}
    \includegraphics[width=0.45\textwidth]{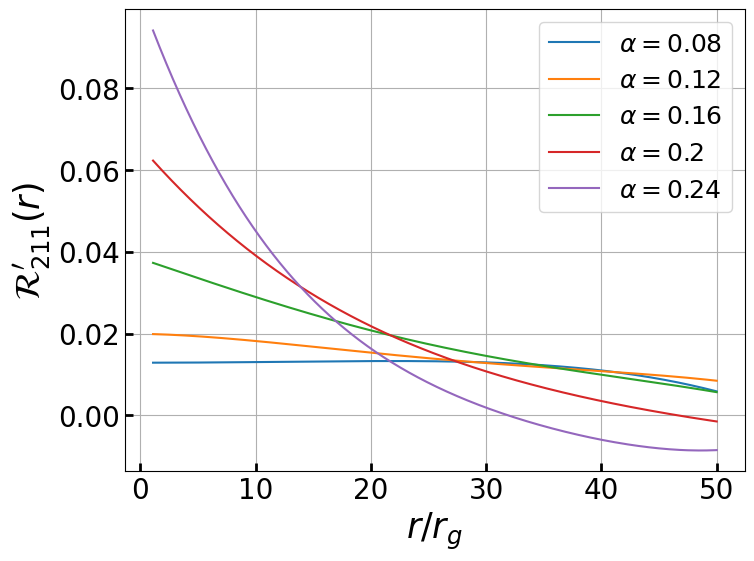}
    \caption{Left: The radial profile $\mathcal{R}_{211}(r)$ of the wavefunction for $a=0.99$ obtained from the numerical result. Right: The derivative of $\mathcal{R}_{211}(r)$ with respect to the radial coordinate $r$.
    }
    \label{Rprofile}
\end{figure}

\section{Photon propagation in the dilaton cloud and plasma}
\label{ppdp}

The coupling function $f(\phi)$ of the dilaton field $\phi$ is non-minimally coupled to the electromagnetic field tensor $F_{\mu\nu}=\partial_{\mu}A_{\nu}-\partial_{\nu}A_{\mu}$ in the following Lagrangian:
\begin{align}
    \mathcal{L}=-\frac{1}{4}f(\phi)F_{\mu\nu}F^{\mu\nu} - e A_{\mu}J^{\mu} ~,
    \label{Larg}
\end{align}
as derived from the low-energy effective action of string theory \cite{Green:1987sp}. We have introduced the electromagnetic four-current $J^\mu$ to account for plasma effects in the accretion environment.

The non-minimal coupling between the dilaton and electromagnetic fields could modify the propagation of photons. To systematically capture these modifications, we employ the geometric optics approximation \cite{thorne2000gravitation} by adopting $f(\phi)=1-g_{\phi\gamma}\phi$ as a benchmark model, where $g_{\phi\gamma}$ is the coupling constant between $\phi$ and $F_{\mu\nu}$. Using the relations $\vec{E}=-\nabla\Phi-\partial_t\vec{A}$ and $\vec{B}=\nabla\times \vec{A}$, the variation of the Lagrangian (\ref{Larg}) with respect to $A^{\mu}=(\Phi, \vec{A})$ yields the following set of modified Maxwell equation:
\begin{align}
& \nabla \cdot \vec{E}=g_{\phi\gamma}(1-g_{\phi\gamma}\phi)^{-1} \nabla \phi \cdot \vec{E} +\rho(1-g_{\phi\gamma}\phi)^{-1} ~, \label{MWE1} 
\\
& \nabla \times \vec{B}-\frac{\partial \vec{E}}{\partial t}=g_{\phi\gamma}(1-g_{\phi\gamma}\phi)^{-1}\left(\nabla \phi \times \vec{B}-\frac{\partial \phi}{\partial t} \vec{E}\right)+\vec{J}(1-g_{\phi\gamma}\phi)^{-1}~, \label{MWE2} 
\\
& \nabla \times \vec{E}+\frac{\partial \vec{B}}{\partial t}=0 ~, \label{MWE3} 
\\
& \nabla \cdot \vec{B}=0 ~,
\end{align}
which also leads to the modified Helmholtz equation:
\begin{align}
    \nabla^2 \vec{E}-\frac{\partial^2 \vec{E}}{\partial t^2}&=g_{\phi\gamma}^2\left(1-g_{\phi\gamma}\phi\right)^{-2}\left( \frac{\partial\phi}{\partial t}\nabla\phi\times\vec{B}-\left( \frac{\partial\phi}{\partial t}\right)^2\vec{E}+\nabla\phi\cdot\left(\nabla\phi\cdot\vec{E}\right) \right) \nonumber \\
    &+g_{\phi\gamma}\left(1-g_{\phi\gamma}\phi\right)^{-1}\left( \nabla\frac{\partial\phi}{\partial t}\times\vec{B}-\frac{\partial^2\phi}{\partial t^2}\vec{E}+\vec{E}\times\left(\nabla\times\nabla\phi\right)+\left(\vec{E}\cdot\nabla\right)\nabla\phi \right) \nonumber \\
    &+\left(1-g_{\phi\gamma}\phi\right)^{-1}\left(\frac{\partial\vec{J}}{\partial t}+\nabla\rho\right)+g_{\phi\gamma}\left(1-
    g_{\phi\gamma}\phi\right)^{-2}\left(\frac{\partial\phi}{\partial t}\vec{J}+\rho\nabla\phi\right) \nonumber \\
    &+g_{\phi\gamma}\left(1-g_{\phi\gamma}\phi\right)^{-1}\left( \left(\nabla\phi\cdot\nabla\right)\vec{E}-\frac{\partial\phi}{\partial t}\frac{\partial\vec{E}}{\partial t} \right) ~.
    \label{Helm}
\end{align}

In the plasma environment, the four-current $J^{\mu}$ in a local frame is decomposed as $J^{\mu} = (\rho, \vec{J})$, where $\rho$ and $\vec{J}$ denote the charge density and current density, respectively. Since the proton mass is approximately three orders of magnitude larger than that of the electron, the collective dynamics of protons can be neglected in most astrophysical plasma environments \cite{landau1987fluid,jackson1998classical}. We therefore consider only the contribution from electrons, i.e., $\rho = e n_e$ and $\vec{J} = e n_e \vec{v}_e$, where $e$, $n_e$, and $\vec{v}_e$ are the electron charge, number density, and velocity, respectively. The current density is driven by the electric field $\vec{E}$, and the dynamics of electrons are governed by the equation $m_e \partial_t \vec{v}_e = e \vec{E}$. The charge and current densities are related through the continuity equation:
\begin{align}
    \frac{\partial \rho}{\partial t}+\nabla \cdot \vec{J}=0 ~.
    \label{cont}
\end{align}
Eqns.~\eqref{Helm} and \eqref{cont} together describe the dynamics of electromagnetic waves interacting with the plasma. It would be almost impossible to obtain analytical solutions by directly solving these equations, since they are highly nonlinear. However, we could adopt the approach of geometric optics to simplify these equations which allows us to gain some useful insights by performing some simple analysis without fully solving them. The ansatz for the Maxwell field in the geometric optics reads
\begin{align}
    \vec{E}=\vec{E}_0 e^{iS},\ \vec{B}=\vec{B}_0 e^{iS} ~,
    \label{solgo}
\end{align}
and the assumption that large gradients of $\vec{E}$, $\vec{B}$ in time and space are described by the real phase $S$, while all other changes are absorbed by a complex, slowly evolving amplitude $\vec{E}_0$ or $\vec{B}_0$. We next examine the characteristic variation scales of relevant physical fields relative to that of $S$. For the dilaton field generated via the superradiance process, we have $\left|\partial_t \phi / \phi\right| \approx \mu$, and $\left|\nabla \phi / \phi\right| \ll \left|\partial_t \phi / \phi\right|$ due to its non-relativistic behavior. Since the Compton wavelength $1/\mu \gtrsim r_g$, as required by the superradiance condition, the phase $S$ associated with photons of wavelength $\lambda \ll r_g$ varies on much shorter temporal and spatial scales than the dilaton field $\phi$, i.e., $\lambda\ll 1/\mu$. Similarly, as plasma dynamics near the black hole are governed primarily by gravity, the electron number density satisfies $\left\vert \nabla n_e / n_e \right\vert \sim 1 / r_g$, implying that the variation scale of $S$ is much smaller than that of the plasma. Consequently, substituting the ansatz Eq.\,(\ref{solgo}) into Eq.\,(\ref{Helm}), and neglecting all derivatives of $\phi$, $n_e$, $\vec{E}_0$, $\vec{B}_0$, the leading-order geometric optics approximation yields
\begin{align}
   -E_j k_\mu k^\mu=\omega_p^2 E_j(1-g_{\phi\gamma}\phi)^{-1} ~,
   \label{maineq}
\end{align}
where we have defined $k_\mu\equiv\partial_\mu S$ and $\omega_p$ is the plasma frequency defined as $\omega_p^2=n_e e^2/m_e$. The existence of nontrivial solutions to Eq.,(\ref{maineq}) requires the vanishing of the operator acting on $\vec{E}$, leading to the dispersion relation $\mathcal{H}=0$ with
\begin{align}
    \mathcal{H}=\vec{k}^2-\omega^2 + \omega_p^2(1-g_{\phi\gamma}\phi)^{-1} ~,
\end{align}
where $\omega\equiv-\dot{S}$, $\vec{k}\equiv\nabla S$. In the Hamiltonian formalism of geometric optics, photon trajectories are identified with worldlines that satisfy the constraint $\mathcal{H}=0$ \cite{breuer1980propagation,thorne2000gravitation}, from which their dynamics follow via the standard Hamiltonian equations of motion. The modified dispersion relation dynamically couples the photon’s effective mass which arises from plasma interactions to the surrounding dilaton field. As a result, temporal oscillations of the dilaton background are expected to induce periodic variations in the effective photon mass, thereby perturbing photon trajectories and imprinting observable signatures on the morphology of the black hole's photon ring.

Gravitational effects can be incorporated by introducing the minimal coupling between the electromagnetic field and the spacetime metric into Eq.\,(\ref{Larg}). At the leading-order geometric optics approximation, derivative terms of the metric in the modified dispersion relation can be neglected, which is equivalent to replacing the flat spacetime metric $\eta_{\mu\nu}$ with the curved spacetime metric $g_{\mu\nu}$, yielding:
\begin{align}
    \mathcal{H}=\frac12\left( g^{\mu\nu}k_\mu k_\nu + \omega_p^2(1-g_{\phi\gamma}\phi)^{-1} \right) ~.
    \label{Hami}
\end{align}
From this we can trace rays via a system of Hamiltonian optics equations
\begin{align}
    \frac{d x^\mu}{d \lambda} = \frac{\partial \mathcal{H}}{\partial k_\mu}, \quad \frac{d k_\mu}{d \lambda} = -\frac{\partial \mathcal{H}}{\partial x^\mu} ~,
    \label{eomphoton}
\end{align}
where $\lambda$ is an arbitrary worldline parameter. In the absence of plasma ($\omega_p=0$), Eq.\,(\ref{eomphoton}) reduces to the null geodesic equation for light propagation in vacuum, implying that the superradiant dilaton field does not affect photon trajectories in this case.

\section{Simulation of the black hole photon ring}
\label{sbr}

\subsection{Ray-tracing algorithm}

In order to extract observational signatures of the superradiant dilaton cloud, we employ the ray-tracing algorithm to simulate the formation of photon ring around the Kerr black hole, which is surrounded by both the dilaton cloud and plasma. The reversibility of light paths in this approach allows us to assume that photons are initially emitted from the observer’s image plane, and then follow trajectories governed by Eq.\,(\ref{eomphoton}) with three types of trajectories: direct, lensed and photon ring. Hence, we are able to determine the location of photon ring in the image plane. The observer's position is specified by $(r_o, \theta_o, \phi_o)$ in the BL spherical coordinates $(r,\theta,\phi)$ of the Kerr spacetime. We assume the observer is located far away from the black hole, where the spacetime can be approximated by the Minkowski metric. In the observer's frame, the coordinate of a point in the image space is labelled as $(X,Y,Z)$, which is related to the Cartesian coordinates $(x,y,z)$ associated with the BL system through the following transformation:
\begin{align}
& x=r_o \sin \theta_o + Z \sin\theta_o - Y \cos \theta_o ~, \nonumber 
\\
& y=X ~, \nonumber 
\\
& z=r_o \cos \theta_o + Z \cos\theta_o + Y \sin \theta_o ~,
\label{trans1}
\end{align}
where we have set $\phi_o=0$ for simplicity. The coordinate transformation between BL Cartesian and spherical coordinate systems is given by
\begin{align}
& r=\sqrt{x^{2}+y^{2}+z^{2}} ~, \nonumber \\
& \theta=\arccos \frac{z}{r} ~, \nonumber \\
& \phi=\arctan \frac{y}{x} ~.
\label{trans2}
\end{align}
Eqs.\,(\ref{trans1}) and (\ref{trans2}) allows that any point $(X,Y,0)$ in the observer's image plane can be mapped to BL spherical coordinates $(r,\theta,\phi)$ of the black hole, serving as the initial position for light rays in the ray-tracing algorithm. To determine the initial condition of momenta $(\dot{r},\dot{\theta},\dot{\phi})$ for light rays, we differentiate Eq.\,(\ref{trans2}) with respect to the affine parameter $\lambda$ to obtain
\begin{align}
    \dot{r}&=\dot{Z} \cos \theta \cos \theta_o-p_0 \sin \theta \sin \theta_o \cos \phi ~, \nonumber \\
    \dot{\theta}&=\frac{\dot{Z}}{r}\left[\sin \theta_o \cos \theta \cos \phi-\sin \theta \cos \theta_o\right] ~, \nonumber \\
    \dot{\phi}&=-\frac{\dot{Z}}{r \sin \theta} \sin \theta_o \sin \phi ~.
\end{align}
Here, we consider emitted light rays perpendicular to the image plane, i.e., $\dot{X}=\dot{Y}=0$. For $\dot{t}$, the Hamiltonian constraint $\mathcal{H}=0$ at spatial infinity provides
\begin{equation}
    \dot{t}=\beta+\sqrt{\beta^2+\gamma} ~,
\end{equation}
where
\begin{equation}
    \beta\equiv -\frac{g_{ti} \dot{x}^i}{g_{tt}} ~, \ \gamma\equiv -\frac{g_{ij} \dot{x}^i\dot{x}^j}{g_{tt}} ~.
\end{equation}
Using the initial conditions specified above, we integrate the differential equations (\ref{eomphoton}) backwards with a second-order Runge-Kutta integrator. The integration proceeds until the photon either falls into the black hole or escapes to infinity.
In order to maintain the numerical precision near the event horizon, we implement a variable step size strategy. At each integration step, we compute the change rates in the direction of $r$, $\theta$, $\phi$ respectively. The final adaptive step size is determined by taking the harmonic mean of these directional rates of change, and constrained by prescribed upper and lower bounds to ensure numerical stability and precision.

\subsection{Results} 

The photon ring of black hole corresponds to the gravitationally lensed image of the photon capture region, where photons occupy unstable orbits with constant radii. In static spherically symmetric spacetimes (e.g., Schwarzschild geometry), this region reduces to a sphere with radius $r=3r_g$. Small perturbations of these unstable photon orbits lead to exponential divergence, resulting in trajectories that either escape to infinity or fall into the black hole \cite{Gralla:2019xty}. Photons that eventually escape to infinity can orbit the black hole multiple times, allowing more photons to accumulate along the same path and thus enhancing the observed intensity. In the case of an optically thin emission disk, the observed intensity at a given image-plane point depends on how many times the associated photon path crosses the disk, giving rise to a characteristic photon ring structure described as:
\begin{align}
    I^{\mathrm{obs}}(X, Y)=\left.\sum_{j=1}^{n} g(r)^4 I(r)\right|_{r=r_j(X, Y)} ~,
\end{align}
where $r_j(X, Y)$ denotes the radial coordinate of the $j^{\mathrm{th}}$ intersection with the disk plane outside the horizon, and $n$ denotes the total number of intersections. $I(r)$ is the emission profile of the disk, and $g(r)$ represents the redshift factor.

\begin{figure}[H]
    \centering
    \includegraphics[width=0.42\textwidth]{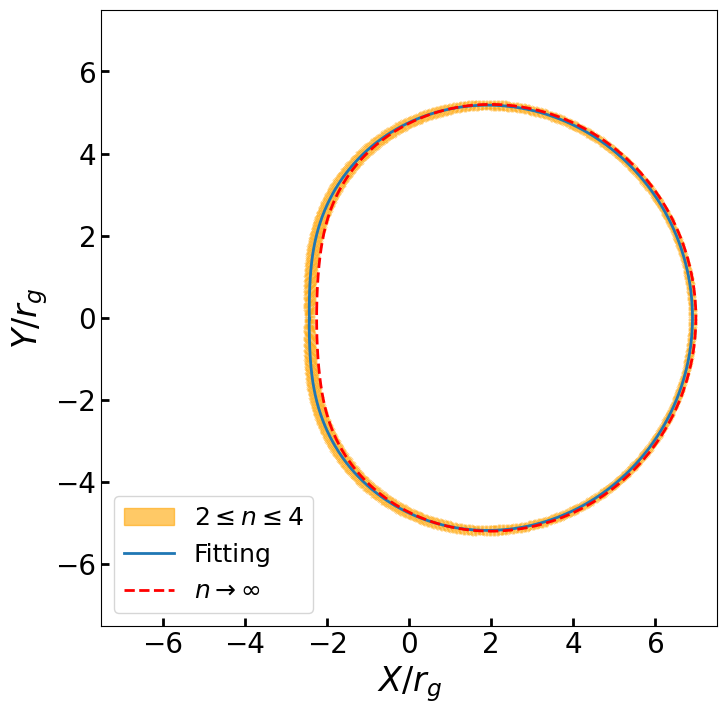}
    \includegraphics[width=0.42\textwidth]{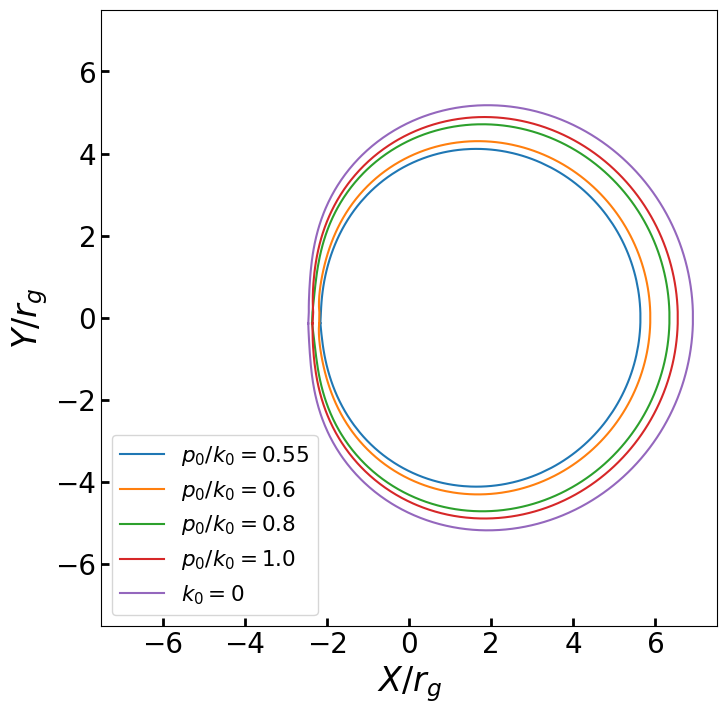}
    \caption{Left: The figure shows the photon ring obtained from the analytical result for $n\to\infty$ (red dashed line), as well as the region enclosed by the photon subrings with $2\leq n\leq 4$ in the case of $k_0=0$, $a=0.99$ and $\theta_o=\pi/2$ (orange). A polynomial fitting method was used to fit this region with a smooth curve (blue), which closely approximates the contour of the $n\to\infty$ photon ring. Right: For $a=0.99$ and $\theta_o=\pi/2$, the photon ring contours formed by different photon frequencies $p_0$, which enters the numerical computation through the ratio $p_0/k_0$.
    }
    \label{Kerr}
\end{figure}

For an extreme Kerr black hole with $a = 0.99$ and $\theta_o = \pi/2$, and neglecting plasma and dilaton coupling effects, the asymptotic photon ring in the limit $n \to \infty$ — derived analytically — is shown as a red dashed curve in the left panel of Fig.\,\ref{Kerr}. In comparison, the orange region represents the envelope formed by photon subrings with winding numbers $2 \leq n \leq 4$, obtained via numerical integration of photon geodesics, while the blue contours correspond to a 10th-order polynomial fit to this region. For the chosen parameters, the analytic $n \to \infty$ photon ring closely traces the outer boundary of the region spanned by the numerically computed subrings. Since a precise numerical determination of the $n \to \infty$ trajectories would demand unattainable computational accuracy, we fix $a = 0.99$ and $\theta_o = \pi/2$ in the subsequent analysis and concentrate on the region defined by subrings with $2 \leq n \leq 4$. This region is effectively characterized using a 10th-order polynomial fit, which enhances computational efficiency and facilitates the detection of subtle distortions introduced by the superradiant dilaton field.

Eq.\,(\ref{Hami}) reveals that the superradiant dilaton field manifests its effects by modifying the effective photon mass induced by the plasma. To examine its impact on the black hole photon ring, a spatial profile for the plasma density must be specified. Here, we adopt a power-law form:
\begin{align}
    \omega_p^2=\frac{k_0^2}{r^h} ~.
    \label{plamod}
\end{align}
The power-law index $h$ characterizes the radial fall-off of the plasma density, and typically depends on the assumed accretion model. For instance, spherical Bondi accretion gives $h=1.5$, while certain disk-like profiles may yield different values \cite{frank2002accretion}. As an example, we fix $h=1$ in the following discussion. In the right panel of Fig.,\ref{Kerr}, the photon ring contours formed by different photon frequencies $p_0/k_0$ are plotted. It can be observed that lower photon frequencies result in a smaller overall size of the photon ring, which is consistent with the conclusions from previous studies \cite{Huang:2018rfn}.

\begin{figure}[H]
    \centering
    \includegraphics[width=0.88\textwidth]{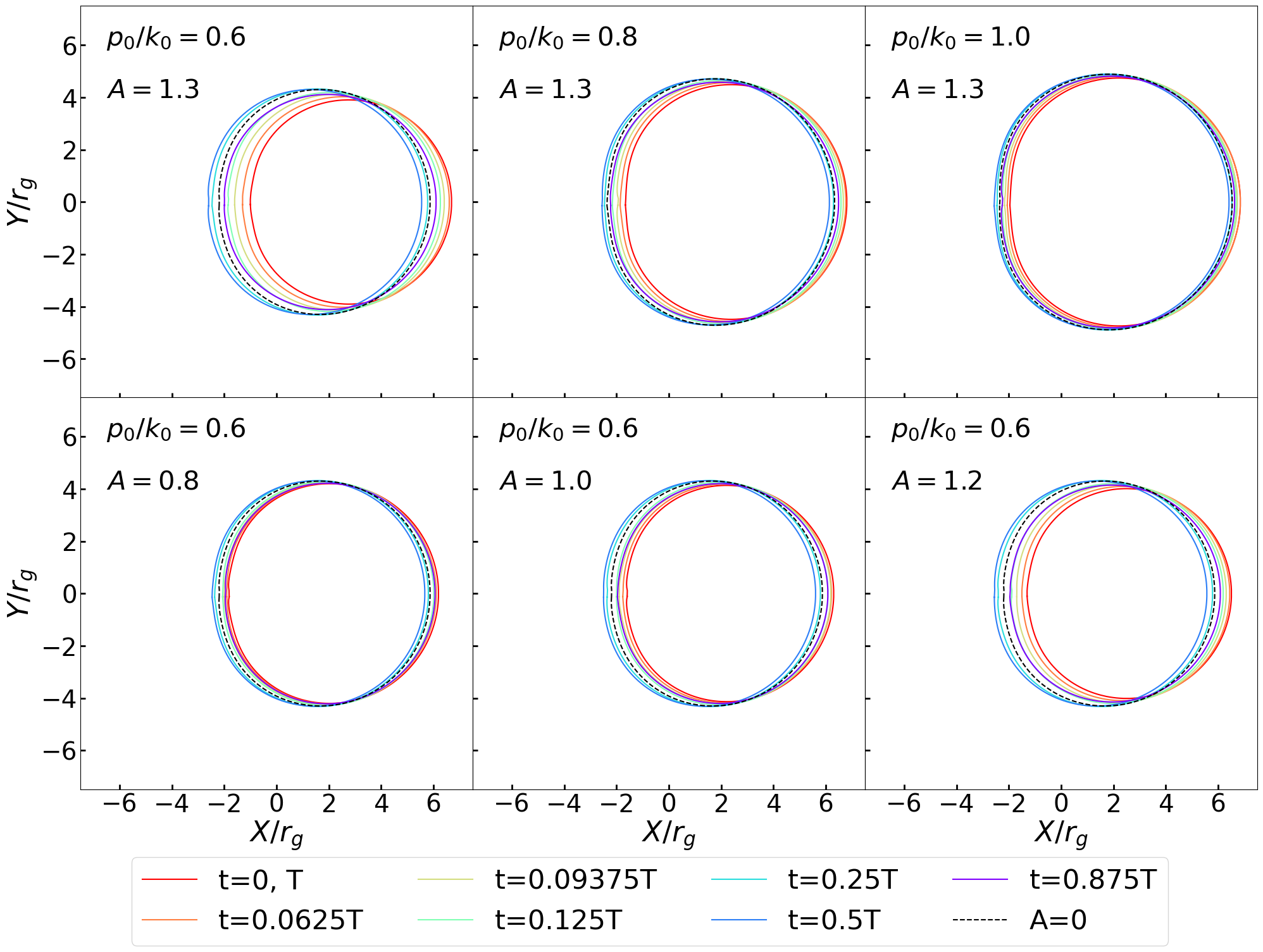}
    \caption{The evolution of the photon ring over time for different values of $p_0/k_0$ and $A=g_{\phi\gamma}\phi_{\text{max}}$, where $T\equiv 2\pi/\alpha$ represents the oscillation period of the superradiant dilaton field. Different rainbow colors represent the temporal sequence of evolution. The black dashed line represents the case without the dilaton field ($A=0$). The black hole spin and inclination are fixed at $a=0.99$, $\theta_0=\pi/2$.
    }
    \label{Ap0}
\end{figure}

Next, we take into account the dilaton field generated by superradiance. Unlike the usual ray-tracing results, the propagation path of photons depends on their arrival time on the image plane, which is caused by the time-oscillating effect of the dilaton field. In Fig.\,\ref{Ap0}, we illustrate the evolution of the photon ring over time for different values of $p_0/k_0$ and $A\equiv g_{\phi\gamma}\phi_{\text{max}}$, with the temporal sequence indicated by rainbow colors. It can be observed that, under the influence of the dilaton field, the shape and size of the photon ring undergo periodic changes, with the period matching the oscillation period $T\equiv 2\pi/\mu$ of the dilaton field. As $p_0/k_0$ decreases, indicating an enhancement of the plasma effect, the amplitude of the photon ring's variation increases. This aligns with the implication of Eq.\,(\ref{Hami}): the dilaton manifests its effect by altering the plasma frequency, and thus the strength of its effect depends on the magnitude of the plasma effects. Moreover, the degree of change in the photon ring’s shape increases with $A$, which is consistent with the intuitive conclusion that a larger dilaton field leads to stronger effects.

\begin{figure}[H]
    \centering
    \includegraphics[width=0.75\textwidth]{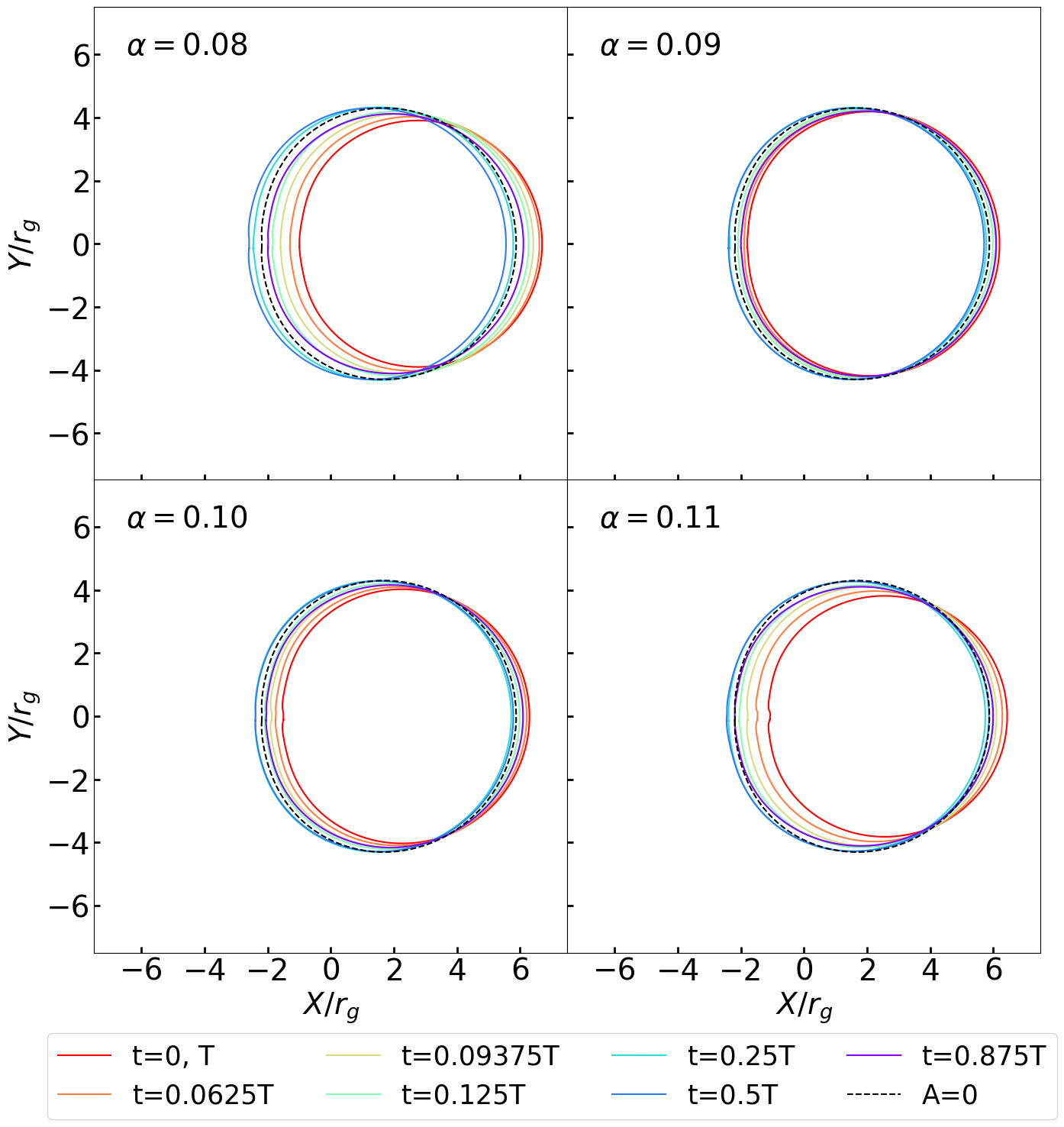}
    \caption{The time evolution of the photon ring for different values of $\alpha$, with $A$ fixed at $1.3$ and $p_0/k_0$ fixed at 0.6. The black hole spin and inclination are fixed at $a=0.99$, $\theta_0=\pi/2$.
    }
    \label{alpha}
\end{figure}

\begin{figure}[H]
    \centering
    \includegraphics[width=0.47\textwidth]{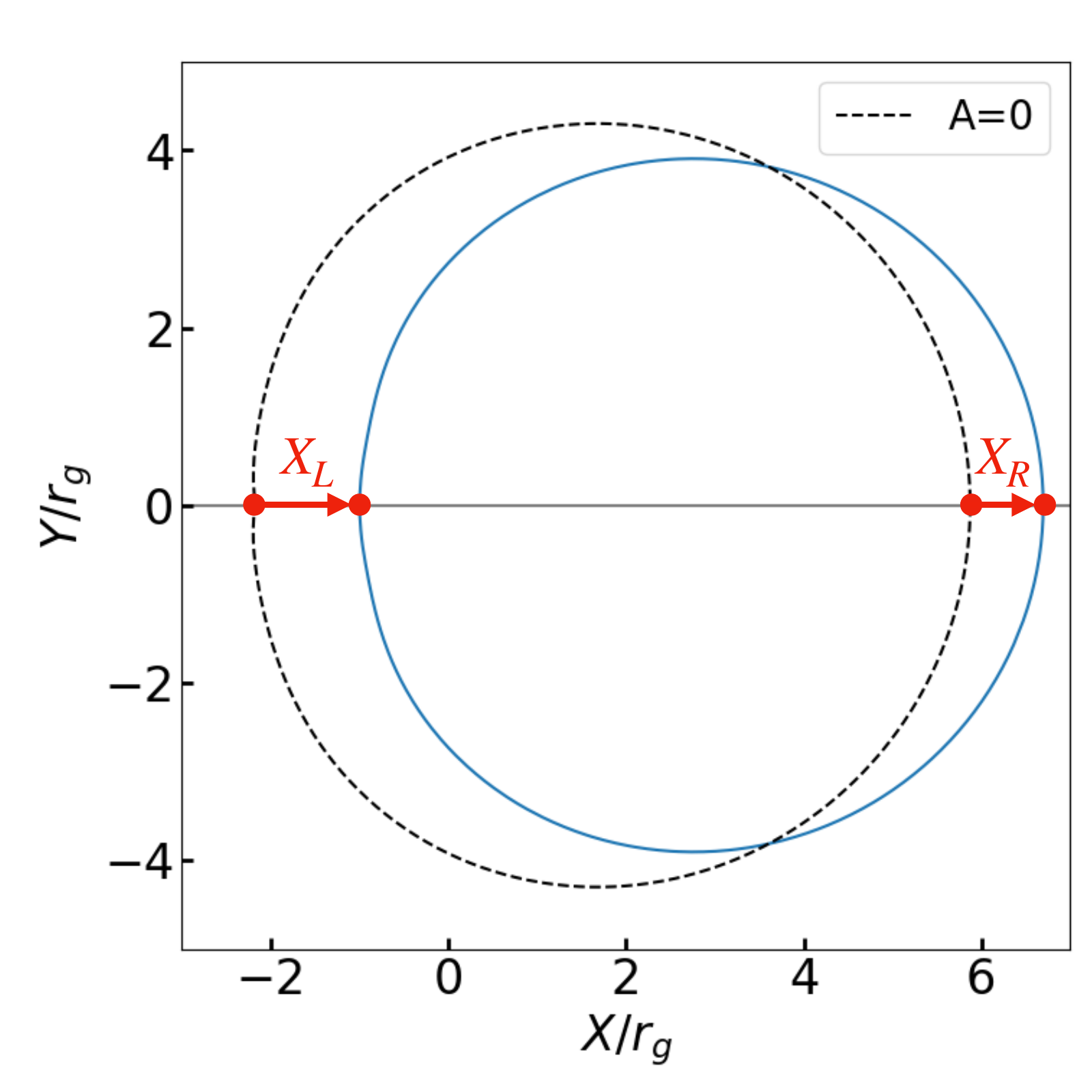}
    \caption{Definition of the photon ring deformation $\delta_\phi$ and $\epsilon_\phi$ under parameters $a=0.99$ and $\theta_o=\pi/2$. The black dashed line shows the photon ring contour in the absence of the dilaton effect ($A = 0$), while the blue line illustrates the contour at a given moment under the influence of the dilaton cloud. $X_L$ and $X_R$ represent the displacements of the left and right endpoints of the ring along the $Y = 0$ line, respectively. The parameter $\delta_\phi = X_R - X_L$ characterizes the variation in the size of the photon ring, whereas $\epsilon_\phi = X_R + X_L$ captures the overall lateral shift of the ring.
    }
    \label{example}
\end{figure}

We then discuss the influence of the dilaton mass on the distortion of the black hole photon ring. In addition to affecting the oscillation period of the dilaton superradiant cloud according to Eq.\,(\ref{phiprofile}), its mass also modifies the radial field profile $\mathcal{R}_{211}(r)$, as shown in Fig.\,\ref{Rprofile}. Both of these effects can influence the propagation of photons via Eq.\,(\ref{eomphoton}), making the analysis significantly complex. In Fig.\,\ref{alpha}, we illustrate the temporal oscillations of the photon ring for different values of $\alpha=\mu r_g$, where we can find that this complexity suggests that the oscillation amplitude of the photon ring exhibits a non-monotonic dependence on $\alpha$. In order to disentangle the impact of varying dilaton masses on the radial profile $\mathcal{R}_{211}(r)$, and thus to identify how the oscillation period of the dilaton field affects the photon ring evolution, we fix $\mathcal{R}_{211}(r)$ to that of $\alpha=0.08$. In this setup, different values of $\alpha$ correspond solely to different oscillation periods of the dilaton cloud. According to Fig.\,\ref{Ap0} and Fig.\,\ref{alpha}, the maximal dilaton-induced deformation of the photon ring occurs along $Y = 0$ on the image plane for the black hole parameters $\theta = 90^\circ$ and $a = 0.99$. Accordingly, the photon ring deformation is quantified by the relative displacements $X_L$ and $X_R$ of its left and right endpoints with respect to case without the dilaton effect, as illustrated in Fig.\,\ref{example}. Based on this, one can define $\delta_\phi = X_R - X_L$ to represent the change in the overall size of the ring, and $\epsilon_\phi = X_R + X_L$ to describe the transverse shift of the ring.

\begin{figure}[H]
    \centering
    \includegraphics[width=0.465\textwidth]{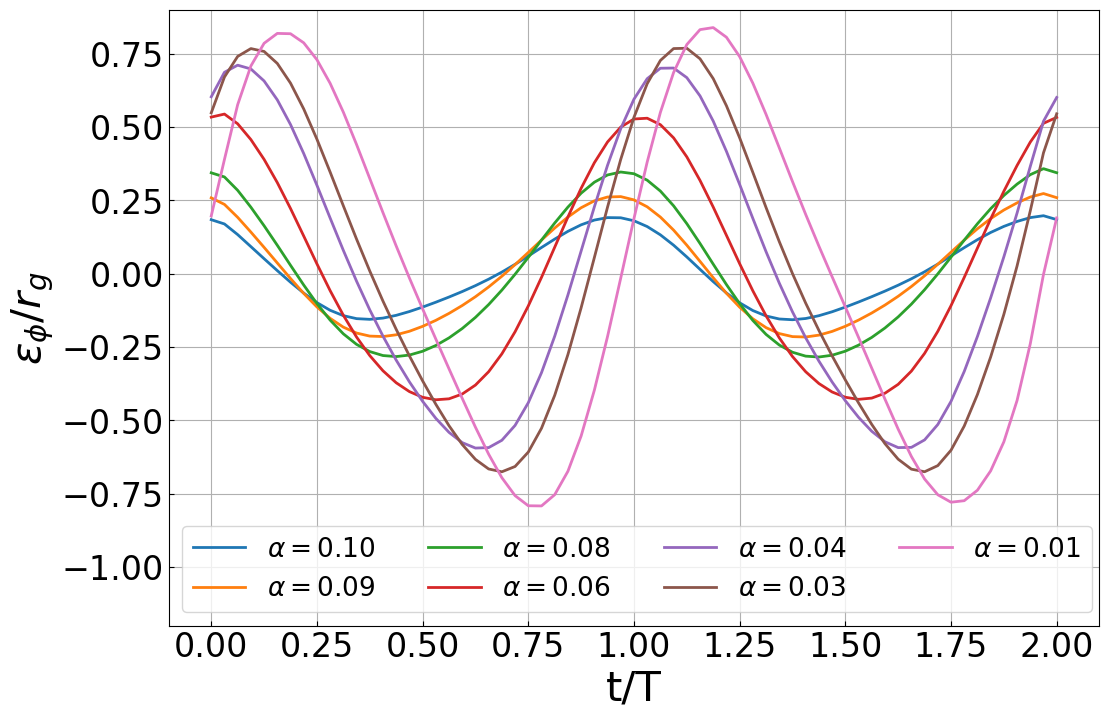}
    \includegraphics[width=0.465\textwidth]{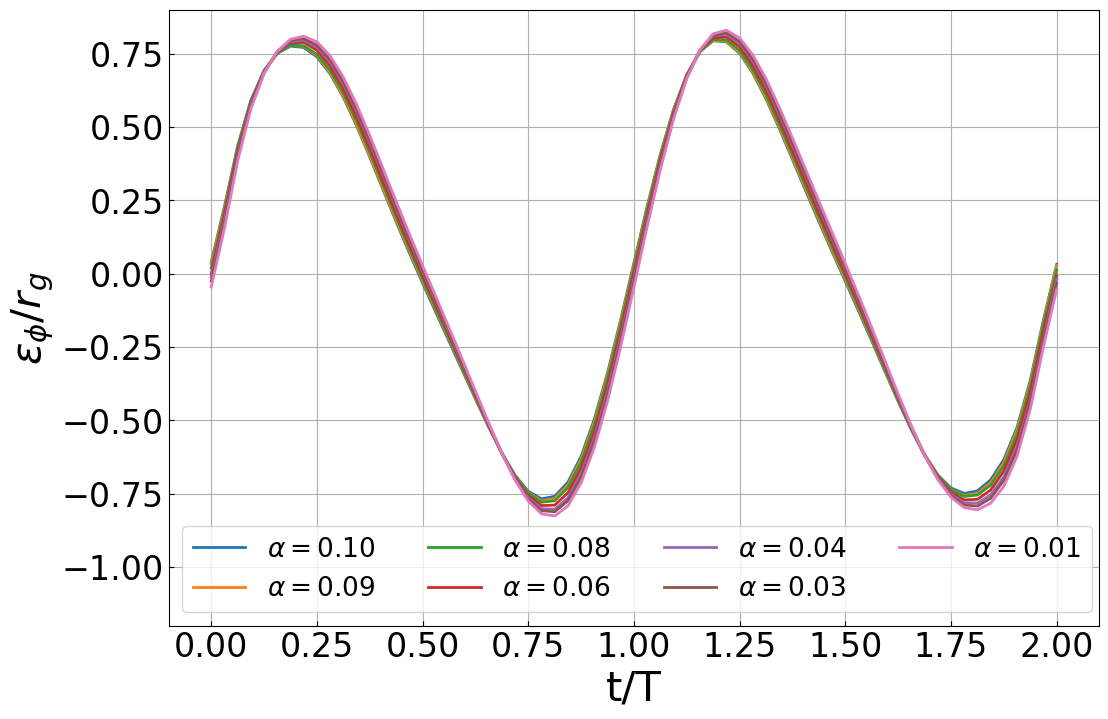}
    \includegraphics[width=0.465\textwidth]{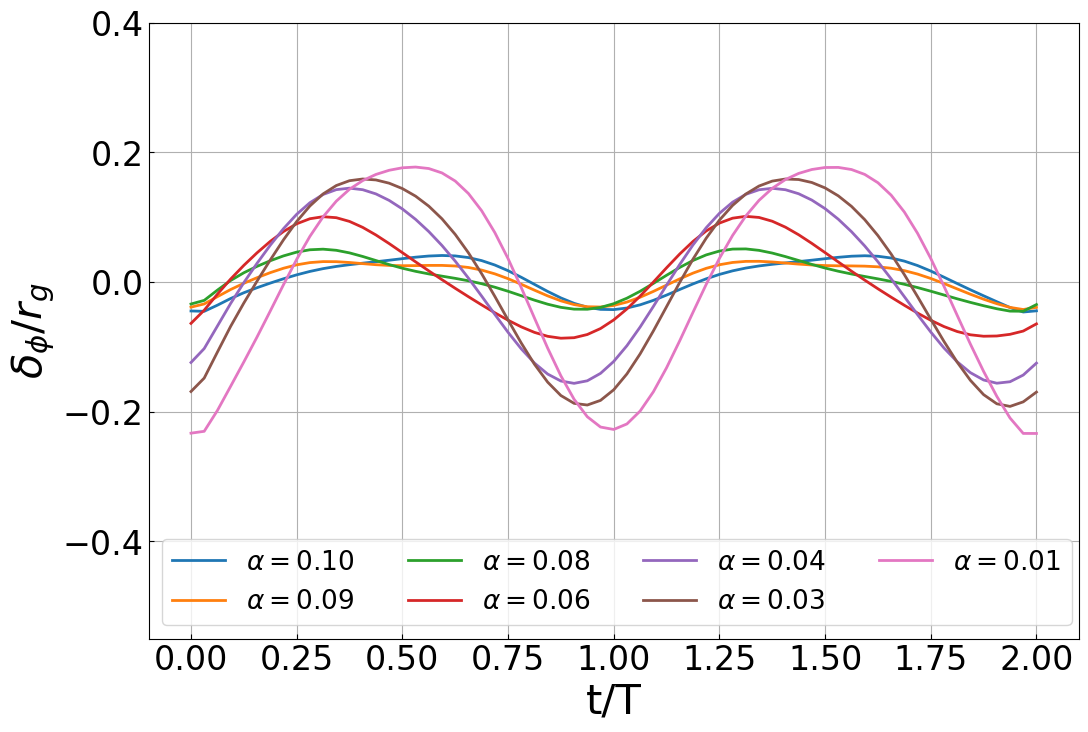}
    \includegraphics[width=0.465\textwidth]{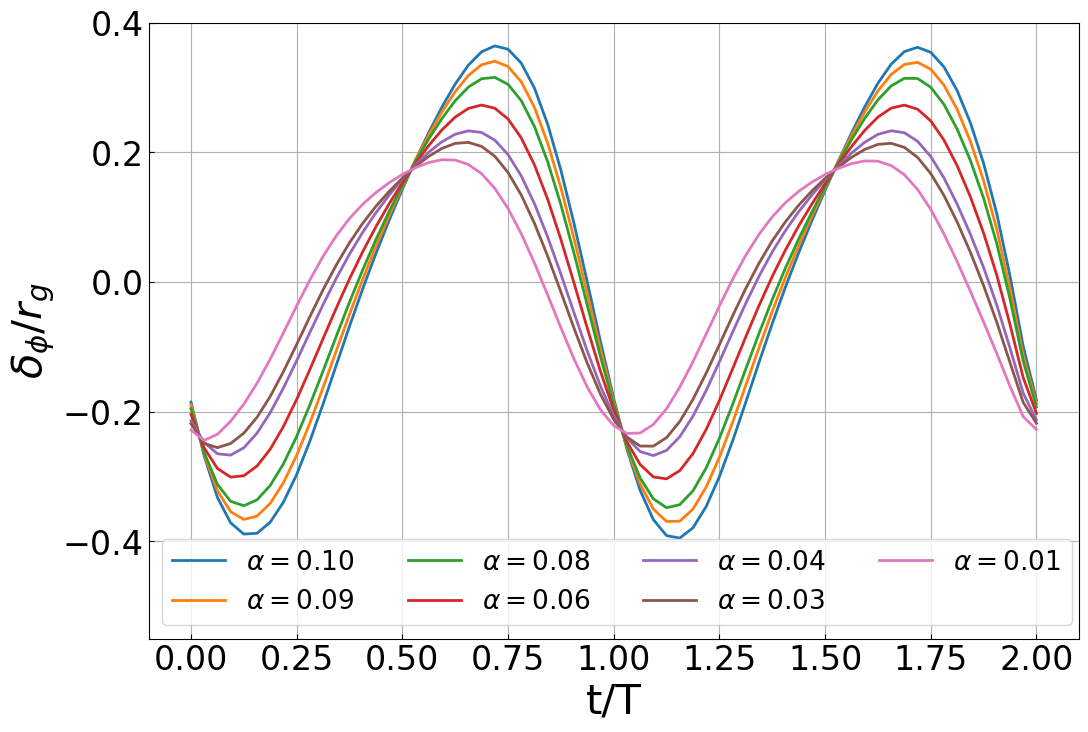}
    \caption{Time evolution of the photon ring deformation $\delta_\phi$ and $\epsilon_\phi$ for different values of $\alpha$, with $A = 0.5$, $p_0/k_0 = 0.6$, $a=0.99$ and $\theta_o=\pi/2$. The radial profile $\mathcal{R}_{211}(r)$ of the dilaton field is fixed to the form corresponding to $\alpha = 0.08$. Horizontal axis is normalized by the oscillation period $T=2\pi/\mu$ of the dilaton field. Left: Evolution of $\delta_\phi$ and $\epsilon_\phi$ with washout effects. Right: Evolution of $\delta_\phi$ and $\epsilon_\phi$ after removing washout effects by choosing the limit $\alpha \to 0$ ($T \gg L$).
    }
    \label{evolu}
\end{figure}

In the left panel of Fig.\,\ref{evolu}, we show the time evolution of the photon ring deformation $\epsilon_\phi$ and $\delta_\phi$ under different values of $\alpha$ over two oscillation periods $T/r_g=2\pi/\alpha$, and other parameters are chosen as $A=0.5$, $p_0/k_0=0.6$. From this figure, we observe that as $\alpha$ increases, the amplitude of the photon ring oscillations decreases. This can be explained by the washout effect of the time-dependent oscillatory terms in the modified geodesic equation (Eq.\,(\ref{eomphoton})) during the integration process. Specifically, if we temporarily neglect the effects of gravity, the correction to the photon velocity $\Delta v$ satisfies $\Delta v\sim\int_{L_p} D(t(\lambda))d\lambda$, where $D$ represents the additional term contributed by the dilaton field in Eq.\,(\ref{Hami}). $L_p$ represents the portion of the photon path affected by the dilaton, which is determined by the range of the plasma distribution given in Eq.\,(\ref{plamod}), and can be estimated as $L_p\sim O(10)r_g$. If the oscillation period of the dilaton field is much shorter than $L_p$, i.e., $T\ll L_p$, the effect of $D(t(\lambda))$ will be washed out during the integration, similar to the case of the superradiant axion cloud’s influence on radiation polarization. In that context, $D(t(\lambda))$ represents the axion-induced correction to the Faraday rotation coefficient of the radiation \cite{Chen:2022oad}. Therefore, as $\alpha$ increases, the oscillation amplitude of the photon ring contour induced by the dilaton field becomes smaller.

To illustrate this point more clearly, we consider the quasi-static case where the influence of the dilaton on the photon propagation is approximately time-independent within the propagation scale $L_p$, and thus the washout effect is effectively removed. This approximation becomes valid in the limit $\alpha \to 0$ ($T\gg L_p$). In this case, the oscillation phase of the photon ring depends solely on the time at which it reaches the image plane. In the right panel of Fig.\,\ref{evolu}, we plot the time evolution of the photon ring deformation parameters $\epsilon_\phi$ and $\delta_\phi$ under this approximation, while for comparison, the left panel shows the corresponding results without adopting this approximation. It can be seen that for the same value of $\alpha$, the oscillation amplitude of the ring is larger when the washout effect is removed. As $\alpha$ decreases, the washout effect becomes weaker due to the larger $T$, and the oscillation amplitudes in both cases tend to converge from the comparison of the two panels. For $\alpha = 0.01$, the two cases are nearly identical. Even when the washout effect is removed via the quasi-static approximation, there are still slight differences in the photon ring oscillations for different values of $\alpha$, as shown in the right panel of Fig.\,\ref{evolu}. This arises from the fact that the correction terms in Eq. (\ref{eomphoton}) contain time derivatives of $t$, which contributes terms that depend directly on the dilaton mass $\mu$ rather than on $\mu t$.

\section{Detectability of the dilaton-electromagnetic coupling}
\label{est}

In this section, based on the photon ring oscillation effects induced by the superradiant dilaton field discussed above, we analyze the constraints that current observational capabilities can place on the dilaton-photon coupling coefficient $g_{\phi\gamma}$. The first step is to determine the plasma distribution surrounding the black hole. We consider spherically symmetric accretion of plasma onto a Schwarzschild black hole, treating the infalling matter as a pressureless perfect fluid that consequently follows geodesic motion \cite{synge1960relativity}. The continuity equation for the stationary accretion then yields
\begin{align}
    4 \pi r^2 \rho(r) u^r(r)=-\dot{M}_A=\mathrm{const},
    \label{contp}
\end{align}
where $\dot{M}_A$ is the stationary mass flux. $u^r(r)$ is the radial component of the 4-velocity for matter falling from rest at infinity:
\begin{align}
    u^r(r)=\frac{d r}{d \tau}=-\sqrt{\frac{2 r_g}{r}}.
\end{align}
Substituting into Eq.\,(\ref{contp}), we obtain the rest-mass density profile:
\begin{align}
    \rho(r)=\frac{\dot{M}_A}{4\sqrt{2}\pi r_g^2} \left(\frac{r_g}{r}\right)^{\frac32}.
\end{align}
Since the accretion flow is electrically neutral, the electron number density equals that of protons: $n_e(r) = \rho(r)/m_p$. Using the plasma frequency definition $\omega_p^2 = n_e e^2 / m_e$, the power-law profile $\omega_p^2 = k_0^2 / (r/r_g)^h$ then takes the form:
\begin{align}
    k_0^2=\frac{e^2 \dot{M}_A}{4\sqrt{2}\pi m_e m_p r_g^2},\ h=1.5.
    \label{k02ex}
\end{align}
The mass accretion rate $\dot{M}_A$ can be estimated from the observed luminosity $L$ of the galactic center via the relation $L\approx \eta \dot{M}_A$, where $\eta$ is a dimensionless coefficient characterizing the accretion efficiency. Its value depends on the accretion model and typically ranges from $10^{-4}$ to $0.1$ \cite{1994ApJ...428L..13N,1995ApJ...452..710N,Bisnovatyi-Kogan:1999yex}.

In the next step, we derive an analytical expression that characterizes the periodic variation in the photon ring size caused by the superradiant dilaton field. For simplicity, we neglect the effect of black hole spin in the photon's equations of motion. Previous studies have shown that the photon ring size is primarily determined by the black hole mass, with variations due to spin and inclination contributing at most 10\% \cite{Johnson:2019ljv}. Therefore, neglecting spin is a justified approximation when focusing solely on the overall size variation, characterized by the parameter $\delta_\phi$ in the above discussion. Moreover, since the numerical results above show that the effect of the superradiant dilaton field is suppressed as $\alpha$ increases, we may therefore focus on the quasi-static regime with $\alpha \lesssim 0.1$ to obtain an upper bound on the strength of the dilaton effect. Under these approximations, the photon ring degenerates into a circular ring, whose radius can be analytically determined from Eq.\,(\ref{eomphoton}) as
\begin{align}
    \frac{d}{r_g}=3\sqrt{3}(1-\delta_p-\delta_{\phi}),\ \ \delta_p = 3^{-h-1}\frac{k_0^2}{\omega_0^2}, \ \ \delta_{\phi} = 3^{-h-1}\frac{k_0^2}{\omega_0^2}g_{\phi\gamma}\bar{\phi},
    \label{supmod}
\end{align}
where the dilaton field is assumed to take a nearly constant value $\bar{\phi}$ within the plasma region \cite{Perlick:2015vta}. As shown in Fig.\,\ref{Rprofile}, this approximation becomes increasingly accurate for smaller values of $\alpha$.

To constrain the dilaton-photon coupling constant $g_{\phi\gamma}$, an estimate of the field amplitude $\bar{\phi}$ is required. As the dilaton cloud grows, superradiance can be quenched either by a violent bosenova process or by entering a saturation phase, both driven by enhanced scalar self-interactions \cite{Yoshino:2012kn,Baryakhtar:2020gao,Omiya:2020vji}. These mechanisms place an upper limit on the achievable field amplitude $\bar{\phi}$. However, if we are only interested in deriving the most stringent constraint that current observations can place on $g_{\phi\gamma}$, it is reasonable to neglect these nonlinear effects and instead consider the maximum possible value of the field. Numerical simulations suggest that up to approximately 10\% of the black hole’s mass can be extracted via superradiant growth of the scalar cloud \cite{Herdeiro:2021znw,East:2017ovw}. We therefore adopt a fiducial estimate of the cloud energy as $M_{\text{cloud}} = 0.1 M$. For simplicity, we regard this energy be approximately localized within a spherical region of radius set by the Compton wavelength $\lambda_c$. The typical energy density of the $\phi$ field can therefore be estimated as $\bar{\rho} \approx M_{\text{cloud}}/((4\pi/3)\lambda_c^3)$. Then according to $\bar{\rho}=\mu^2\bar{\phi}^2$, the corresponding field value $\bar{\phi}$ can be estimated as $\bar{\phi}\approx 10^{25}\text{eV}(\alpha/0.01)^{1/2}$.

Finally, the photon ring distortion $\delta_\phi$ in Eq.\,(\ref{supmod}) can be translated into an angular deviation $\delta\beta$ via $\beta = \delta_\phi / r_o$, where $r_o$ is the distance to the black hole. Given the current observational capabilities, the Event Horizon Telescope (EHT), operating at $\lambda = 1.3\,\mathrm{mm}$ with an effective aperture of $D \approx 1.3 \times 10^4\,\mathrm{km}$, achieves an angular resolution of $\delta\beta \approx \lambda / D \approx 20\,\mu\mathrm{as}$ \cite{EHT2019dse}, providing a benchmark for detectable distortions. Incorporating the above analysis, the resulting constraint on the dilaton-photon coupling constant $g_{\phi\gamma}$ is given by
\begin{align}
    g_{\phi\gamma} &\approx 10^{-11}\,\text{GeV}^{-1} \left( \frac{\beta}{20\,\mu\text{as}} \right) \left( \frac{\eta}{10^{-4}} \right) \left( \frac{\lambda}{1.3\,\text{mm}} \right)^{-2} \left( \frac{M}{4.3\times10^6M_\odot}\right) \left( \frac{L}{10^6\,\mathrm{L}_\odot} \right)^{-1} \nonumber \\
    &\times \left( \frac{r_o}{8.3\,\text{kpc}} \right) \left( \frac{\alpha}{0.1} \right)^{-\frac{1}{2}}.
\end{align}
In the above estimate, the black hole mass $M$, luminosity $L$, and distance $r_o$ are chosen as those of $\mathrm{Sgr\,A}^\star$ \cite{EventHorizonTelescope:2022wkp,2008ApJ...689.1044G,1992ApJ...387..189D}, with $\alpha = 0.1$ corresponding to the dilaton mass $\mu = 3.1 \times 10^{-18}\,\mathrm{eV}$. For comparison, the parameters for $\mathrm{M87}^\star$ are $L = 1.8 \times 10^7 L_{\odot}$, $M = 6.5 \times 10^9 M_{\odot}$ and $r_o = 18\,\mathrm{Mpc}$ \cite{EventHorizonTelescope:2019ggy,Gebhardt_2009,Ho_2009}, which yield a weaker constraint on the coupling constant: $g_{\phi\gamma} \approx 8.9 \times 10^{-7}\,\mathrm{GeV}^{-1}$ at $\mu = 4.5 \times 10^{-21}\,\mathrm{eV}$ for $\alpha=0.1$.

In the mass range of $10^{-21} \sim 10^{-17}\,\mathrm{eV}$, the most stringent existing constraints on the dilaton-photon coupling constant $g_{\phi\gamma}$ come from atomic spectroscopy. These experiments, primarily utilizing atomic clocks and their arrays, aim to detect variations in the fine-structure constant induced by ultralight scalar dark matter, and have placed bounds on $g_{\phi\gamma}$ at the level of $10^{-27}\,\mathrm{eV}^{-1}$ \cite{VanTilburg:2015oza,Hees:2016gop,Barontini:2021mvu,Sherrill:2023zah,Filzinger:2023zrs}. In addition, tests of equivalence principle and searches of fifth-force mediated by scalar fields constrain $g_{\phi\gamma}$ at the level of $10^{-21}\,\mathrm{eV}^{-1}$ \cite{Hees:2018fpg,Adelberger:2003zx,fischbach1996ten,KONOPLIV2011401}. Although the current observational sensitivity achievable via superradiance around supermassive black holes yields much weaker bounds, such studies remain valuable given that the nature and production of the dilaton field in superradiance differ fundamentally from those associated with dark matter or fifth-force scenarios.

\section{Conclusion}
\label{con}
In this work, we investigated the impact of black hole superradiance-induced dilaton clouds on photon rings, focusing on the interaction between the dilaton field and electromagnetic fields. Within the geometric optics approximation, we found that the dilaton-photon coupling modifies the plasma frequency, effectively inducing a periodic modulation in the photon’s effective mass. Incorporating this correction into ray-tracing simulations in the black hole spacetime, we demonstrated that the photon ring undergoes periodic oscillations with a period matching that of the dilaton field oscillation.

We numerically examined how the amplitude of these ring oscillations depends on various plasma and dilaton parameters. Our results first confirmed that the oscillation amplitude increases with stronger plasma effects, higher dilaton energy density, and larger photon-dilaton coupling strength. However, its dependence on the dilaton mass is more intricate, as the mass affects both the radial profile of the field and its temporal oscillation. To isolate the impact of the latter one, we fix the radial profile of the dilaton field and only vary its oscillation period. The results show that when the dilaton’s oscillation period $T/r_g=2\pi/\alpha$ is much shorter than the characteristic photon propagation timescale $L_p$ across the plasma region, the photon ring deformation becomes significantly suppressed, which is caused by the washout effect of the path integral over the dilaton-induced correction term.

Finally, based on a spherically symmetric pressureless fluid accretion model, we assessed the detectability of such dilaton-induced photon ring oscillations under current radio observational capabilities. Our estimates suggest that this method can potentially constrain the dilaton-photon coupling constant to $g_{\phi\gamma}\lesssim 10^{-11}\,\text{GeV}^{-1}$ for dilaton masses $\mu \lesssim 10^{-18}\,\mathrm{eV}$.

\acknowledgments

We are grateful for useful discussions with Changhong Li and Yue Zhao, which inspired the key ideas presented in this manuscript. CL is supported by National Natural Science Foundation of China (No. 12447128) and the China Postdoctoral Science Foundation (No. 2024M760720). CC is supported by National Natural Science Foundation of China (No. 12433002) and Start-up Funds for Doctoral Talents of Jiangsu University of Science and Technology. CC thanks the supports from The Asia Pacific Center for Theoretical Physics, The Center for Theoretical Physics of the Universe at IBS, The COSPA Group at the USTC during his visits. XYC is supported by the starting grant of Jiangsu University of Science and Technology (JUST).
All numerics are operated on the computer clusters in the cosmology group at IAS, HKUST.


\bibliographystyle{JHEP}
\bibliography{main.bib}


\end{document}